# Bayesian joint modelling of longitudinal biomarkers to enable extrapolation of overall survival: an application using larotrectinib trial clinical data

## Authors

Louise Linsell, DPhil MSc BSc, louise.linsell@visibleanalytics.co.uk, Visible Analytics, Oxford, UK. [joint first author]

Noman Paracha, MSc BSc, noman.paracha@bayer.com, Bayer Pharmaceuticals, Basel, Switzerland. [joint first author]

Jamie Grossman, PhD MBA BA, jamie.grossman@bayer.com, Bayer Pharmaceuticals, Ohio, USA.

Carsten Bokemeyer, MD, c.bokemeyer@uke.de, University Medical Centre Hamburg Eppendorf, Hamburg, Germany.

Jesus Garcia-Foncillas, MD PhD, jesus.garciafoncillas@oncohealth.eu, University Cancer Institute and the Department of Oncology, University Hospital Fundacion Jimenez Diaz, Autonomous University, Madrid, Spain.

Antoine Italiano, MD PhD, a.italiano@bordeaux.unicancer.fr, Institut Bergonié, Bordeaux, France.

Gilles Vassal, MD PhD, gilles.vassal@gustaveroussy.fr, Gustave Roussy Comprehensive Cancer Center, Villejuif, France.

Yuxian Chen, MSc BSc, yuxian.chen@visibleanalytics.co.uk, Visible Analytics, Oxford, UK.

Barbara Torlinska, PhD MSc MA, barbara.torlinska@visibleanalytics.co.uk, Visible Analytics, Oxford, UK.

Keith R Abrams, PhD MSc BSc, keith.abrams@warwick.ac.uk, Department of Statistics, University of Warwick, UK.

## Corresponding author

Louise Linsell, Visible Analytics, First Floor, 1 Des Roches Square, Witan Way, Witney

OX28 4BE, UK. Email: louise.linsell@visibleanalytics.co.uk.

**Abstract**

***Objectives***

To investigate the use of a Bayesian joint modelling approach to predict overall survival (OS) from immature clinical trial data using an intermediate biomarker. To compare the results with a typical parametric approach of extrapolation and observed survival from a later datacut.

***Methods***

Data were pooled from three phase I/II open-label trials evaluating larotrectinib in 196 patients with neurotrophic tyrosine receptor kinase fusion-positive (*NTRK+*) solid tumours followed up until July 2021. Bayesian joint modelling was used to obtain patient-specific predictions of OS using individual-level sum of diameter of target lesions (SLD) profiles up to the time at which the patient died or was censored. Overall and tumour site-specific estimates were produced, assuming a common, exchangeable, or independent association structure across tumour sites.

***Results***

The overall risk of mortality was 9% higher per 10mm increase in SLD (HR 1.09, 95% CrI 1.05 to 1.14) for all tumour sites combined. Tumour-specific point estimates of restricted mean , median and landmark survival were more similar across models for larger tumour groups, compared to smaller tumour groups. In general, parameters were estimated with more certainty compared to a standard Weibull model and were aligned with the more recent datacut.

***Conclusions***

Joint modelling using intermediate outcomes such as tumour burden can offer an alternative approach to traditional survival modelling and may improve survival predictions from limited follow-up data. This approach allows complex hierarchical data structures, such as patients nested within tumour types, and can also incorporate multiple longitudinal biomarkers in a multivariate modelling framework.

## Main body

### Introduction

Clinical studies are often interested in the association between a clinical biomarker that is measured repeatedly over time and a time-to-event outcome, such as death or disease progression, typically to investigate how changes in the biomarker can be used to predict the occurrence of the event. Joint modelling is a rapidly developing area of biostatistical research that models both the longitudinal and time-to-event data simultaneously to estimate this association (1, 2). It has several potential benefits over more traditional approaches that directly include the biomarker as a covariate in the survival regression model as it can effectively adjust the estimation of biomarker profiles in the presence of informative drop-out, e.g. due to death, which helps to reduce the potential bias introduced by missing data or measurement errors (3). Joint models can predict the time-to-event outcome for censored patients conditional upon their longitudinal biomarker profile up to a specific time point, adjusted for other covariates as appropriate. Patient-specific predictions of survival outcome (for censored patients) can be estimated by extrapolating the results beyond the duration of the trial with associated uncertainty.

Extensions to this include multivariate joint modeling which allows for more than one longitudinal biomarker and/or time-to-event outcome to be fitted simultaneously (4, 5). In addition, hierarchical modelling can be used to explore more complex data structures such as clustering factors above the patient-level clustering of repeated (biomarker) measures over time, such as clinic, study or tumour type, and clustering factors below the patient-level, such as lesions (3).

With the proliferation of histology-independent oncology treatments based on common biomarkers rather than the primary tumour site, and the increased use of basket trials to evaluate them, novel and robust statistical methods are required. The potential for heterogeneity in treatment effect across tumour types and patient clinical characteristics and prognosis will be a growing issue for healthcare appraisals and marketing authorization (6). A surrogate-based modeling approach such as joint modelling, within a Bayesian hierarchical framework, can explicitly model such heterogeneity and provide pooled treatment effects for each tumour type by borrowing information across baskets (tumour types) (3, 7, 8).

This paper demonstrates the use of a Bayesian joint modelling approach to predict overall survival from clinical trial data with incomplete follow-up for all patients. Neurotrophic tyrosine receptor kinase fusion-positive (*NTRK*+) in solid tumours has been shown to be a predictive biomarker for targeted inhibition (9) and clinical trials evaluating the effect of innovative precision therapies on survival are currently underway. However, such trials take many years to complete and robust methods for extrapolating survival from immature follow-up data are required. Using clinical trial data evaluating treatment with larotrectinib in patients with *NTRK*+ tumours (10), we apply a Bayesian joint modelling approach to establish the association between tumour burden and overall survival (OS) to predict future conditional survival. We also provide tumour-specific estimates of OS, using hierarchical joint modelling in which patients are nested within tumour type, and an exchangeable joint modelling approach in which the estimation of the association between OS and SLD for a specific tumour type "borrows strength" from that estimated in other tumour types. We compare the results from the different joint models with a typical parametric approach of extrapolation (11) and with observed survival times from a later cut of the dataset.

## Methods

### *Data*

Survival analysis was based on prospective data from 3 phase I/II open-label trials evaluating the safety and efficacy of larotrectinib in adults and children with neurotrophic tyrosine receptor kinase fusion-positive (*NTRK*+) solid tumours. A phase I dose escalation study (20288, NCT02122913) was completed in adult patients with solid tumours, with an expansion phase ongoing in patients with *NTRK*+ tumours. A phase I/II study in a children and young adults aged 21 years or under is currently underway (SCOUT, NCT02637687), in addition to a phase II basket trial across multiple primary tumour types in adults and adolescents aged 12 years and above (NAVIGATE, NCT02576431). Preliminary efficacy analysis of the integrated dataset suggest that larotrectinib is effective at either shrinking tumours or delaying progression across a range of different cancer types, however median OS was not reached after median follow-up of 38 months (10).

In a datacut taken on 20 July 2021, 244 patients were included in the integrated dataset. One patient with no tumour measurements and 47 children with infantile fibrosarcoma, which has a different disease progression and prognosis, were excluded. The remaining 196 patients were included in the analysis; 13 from the Phase I 20288 trial, 41 from SCOUT and 142 from NAVIGATE. The primary endpoint was OS from the first dose of larotrectinib, and the intermediate biomarker used in the joint models was investigator-assessed sum of diameters of target lesions (SLD). For each patient, the sum of the longest diameter (mm) was taken for between 1 and 5 target lesions measurable according to the Response Evaluation Criteria in Solid Tumours (RECIST) version 1.1 criteria (12). Measurements were taken at every other 28-day cycle during the first year of trial entry and every 3 months thereafter.

***Statistical analysis***

The joint model consisted of a survival submodel for OS linked by an association parameter to a longitudinal submodel for the SLD biomarker. A generic form of the longitudinal sub-model can be written as:

$$y_i(t) = \boldsymbol{X}_i^T(t)\,\beta + \boldsymbol{Z}_i^T(t)\boldsymbol{b}_i + \epsilon_i(t) \quad \epsilon_i(t) \sim N(0,\sigma^2) \qquad [1]$$

$$y_i(t) = m_i(t) + \epsilon_i(t) \qquad [2]$$

where for subject $i$, $y_i(t)$ is the outcome (biomarker) at time $t$, $\boldsymbol{X}_i(t)$ is the design matrix for the fixed effects, $\beta$ are the fixed effects parameters, $\boldsymbol{Z}_i(t)$ is the design matrix for the random effects, and $\boldsymbol{b}_i$ are the random effects and assumed to have Multivariate Normal distribution, i.e. $b_i \sim N[0,\Sigma]$. Equation [1] can be re-written as in [2] with $m_i(t)$ which is referred to as the trajectory function, i.e. the true unobserved value of the biomarker for the $i$th patient at time $t$. A generic form of a proportional hazards survival sub-model can be expressed as:

$$h_i(t) = h_0(t)exp\,(\phi^T\boldsymbol{v}_i) \qquad [3]$$

Where $h_i(t)$ is the hazard at time $t$ for patient $i$, $h_0(t)$is a baseline hazard function, e.g. Weibull, $\phi$ are regression parameters, e.g. log hazard ratios, and $\boldsymbol{v}_i$ is the design matrix for fixed effects. If we define $M_i(t) = \{m_i(s), 0 \leq s \leq t\}$ to be the true unobserved longitudinal profile up to time $t$, then the joint model can then be parameterized as:

$$h_i(t|M_i(t), \boldsymbol{v}_i) = h_0(t)\ exp\left(\phi^T\ \boldsymbol{v}_i + \alpha\, m_i(t)\right) \qquad [4]$$

where,

$$m_i(t) = \boldsymbol{X}_i^T(t)\,\beta + \boldsymbol{Z}_i^T(t)\boldsymbol{b}_i \qquad [5]$$

$m_i(t)$ is termed the current value parameterisation and $\alpha$ is the association parameter, representing the log hazard ratio (HR) for a 1-unit increase the value of the biomarker. Slope association, i.e. rate of change in SLD was also considered as a potential association parameter linking the 2 sub-models.

Three joint models were fitted with different assumptions about the association parameter in relation to tumour type. Firstly, a joint model with a common association parameter was fitted, which assumes that the relationship between SLD and OS is the same for each tumour type, i.e. common $\alpha$ for all tumour types as in equation [4]. The second joint model used an assumption of exchangeability to estimate a separate association parameter for each tumour type from a distribution with a common overall association:

$$h_i(t|M_i(t), \boldsymbol{v}_i) = h_0(t)\exp\left(\phi^T\boldsymbol{v}_i + \alpha_{k_i} m_i(t)\right) \qquad [6]$$

where $\alpha_{k_i}$ indicates the association parameter for the tumour type of the *i*th patient, and

$$\alpha_k \sim N[\alpha, \tau^2] \quad k = 1, \ldots, K \qquad [7]$$

.

i.e. $\alpha$s assumed to be exchangeable across $K$ tumour types, and drawn from a common distribution in which $\alpha$ represents an overall association and $\tau^2$ is the between-type association variance. The third joint model assumed an independent association parameter $\alpha$ for each tumour type, i.e. $\alpha_k$ $k = 1, \ldots, K$ of $K$ possible types, providing a unique estimate for each.

Bayesian bespoke joint models were fitted with parameter values estimated using Markov chain Monte Carlo as implemented in **JAGS** (Just Another Gibbs Sampler) version 4.3.1 (13) using the **rjags** package(14) in **R** version 4.2.0. Models were run with a burn-in period of 50,000 iterations using 3 chains with different initial values, and a further 150,000 iterations used for estimation. Convergence was assessed using the $\hat{R}$ statistic, traceplots and Brooks-Gelman-Rubin statistics (16). The accuracy of estimation was judged sufficient if the Monte Carlo standard error divided by posterior standard deviation for all parameters was less than 5%. Comparative model fit was compared using the Deviance Information Criterion (DIC) (17). For initial and exploratory assessment of the data and a joint modelling approach the **JMBayes2** package in **R** was also used (18).

For the longitudinal biomarker model, vague Uniform prior distributions were used for the population intercept – Uniform (0,60), the population slope – Uniform (0,1) and the between-patient standard deviation – Uniform (0,20). A weakly informative Half-Normal ($0.5^2$) prior distribution was used for between-tumour standard deviation. For the survival model, a parametric Weibull proportional hazards model was used as it had the best statistical fit, and was also the base case distribution used for the larotrectinib National Institute for Health and Care Excellence health technology assessment (NICE HTA) submission (19). A vague Normal ($0,1000^2$) prior distribution was used for the regression coefficients and scale parameter, and a weakly informative Exponential (0.003) prior distribution for the shape parameter. Site of primary tumour was categorised as soft tissue sarcoma, thyroid (differentiated), salivary gland, lung, and other. There were 20 tumour types with less than 20 patients which formed the "other" category.

For each Bayesian joint model, the association between SLD and OS was reported as an association hazard ratio (HR) with 95% credible interval (CrI). Extrapolation from the joint models was then used to obtain patient-specific predictions of OS (by sampling from the posterior predictive distribution of OS for censored patients), conditional on the individual-level biomarker profiles up to the time at which the patient died or was censored, adjusting for informative dropout. These were summarised using restricted mean survival time (RMST) over a 100-year lifespan, median survival, and landmark survival at 10 years with 95% Crls and compared to predictions from a standard parametric Weibull model. RMST at 5 years was compared to the observed RMST from a more recent datacut (July 2023). As part of the validation exercise, we also compared RMST at 5 years to a later datacut taken in July 2023 when patients had been followed for an additional 2 years.

**Results**

Median follow-up for OS was 32.4 months (range 0.4 to 71) for the 196 patients analysed and 58 deaths (29.6%) were observed. The Kaplan-Meier plot for OS is shown in Figure S1; median OS was not estimable. Table S1 summarises the number of patients and mortality events for key baseline clinical covariates. The risk of mortality increased with age and was greater in patients with higher ECOG scores and metastatic spread at treatment initiation. The event rate varied from 20% in soft tissue sarcoma (which included 35 of the 40 pediatric patients) to 47% among tumour types grouped as “other”. Figure 1 displays the SLD trajectories by mortality status, which have a lower and flatter distribution in the group of patients who were censored at their latest visit.

The association parameters and model fit statistics for the three joint models of OS and SLD using a current value association structure with common, exchangeable, and independent assumptions across tumour sites are shown in Table 1. Using the slope association instead of current value resulted in a poorer model fit for the common association model. In this model, the overall risk of mortality was 9% higher per 10mm increase in SLD (HR 1.09, 95% Crl 1.05 to 1.14) for all tumour sites combined. The exchangeable joint model had the best fit (lowest DIC). The tumour-specific association parameters from both the exchangeable and independent joint models varied slightly, suggesting that the strength of association between SLD and OS may differ across tumour site, but further data are required to estimate this with more certainty.

Predictions of OS from treatment initiation over a life expectancy (100 years) for the overall cohort and tumour-specific subgroups using the standard parametric Weibull model and the 3 joint models are presented in Table 2. For the overall cohort, the restricted mean survival time was 9.35 years (95% credible interval (CrI): 5.59 to 14.58) in the exchangeable joint model, compared to 8.04 years (95% CrI: 4.88 to 13.44) in the standard Weibull model. 10-year OS predictions from treatment initiation were also similar; 28.1% (95% Crl: 17.3% to 38.3%) in the exchangeable joint model compared to 27.7% (95% Crl: 13.7% to 39.8%) in the standard Weibull model. For the overall cohort, point estimates for each parameter were generally similar across all models, but higher in the independent joint model. RMST was estimated with less certainty (wider credible intervals), and median survival and 10-year landmark survival was estimated with more certainty (narrower credible intervals) in the joint models compared to the standard Weibull model.

Predictions for each tumour site were more variable across models, due to the small numbers of patients and events within each group. Tumour-specific point estimates of RMST, median and landmark survival were more similar across models for the larger tumour groups (i.e. soft tissue sarcoma and other) versus smaller tumour groups (i.e. lung, salivary gland, and thyroid), and generally estimated with more certainty by the joint models. For the smaller tumour groups, the point estimates for the joint models were similar to each other but markedly different from the standard Weibull model and estimated with greater uncertainty.

Figure 2 displays the extrapolated OS curves from the Weibull proportional hazards model and exchangeable joint model for the overall cohort and by tumour site, demonstrating a poorer prognosis for lung and other tumour site, and a better prognosis for soft tissue sarcoma and thyroid. Estimates for RMST at 5 years from the exchangeable joint model and Weibull proportional hazards model compared to those observed in the later datacut of July 2023 are shown in Table 3. A further 13 patients across the different tumour groups died between the two datacuts. The point estimates were similar across models but estimated with more certainty by the joint model (narrower credible intervals).

**Discussion**

This work builds upon the large body of literature on extrapolating survival data (11), and whilst the use of dynamic tumour growth modelling has been advocated in oncology drug development (20), there has been relatively few examples of adopting a joint modelling approach (21) to facilitate patient-level predictions (conditional on tumour dynamics) in order to extrapolate either overall or tumour site specific survival (22). While a number of mainstream software packages have been developed to implement joint modelling, there has been less development of methods to assess covariate selection and model performance, and to date the application of joint modelling in a HTA context has been limited. This case-study provides an example of the potential application of joint modelling in this context, though our analysis was limited by a small and heterogeneous sample with a low event rate, due to the rare nature of *NTRK* gene fusion driven cancer.

A number of modelling assumptions and decisions were made in the formulation of the joint models. Whilst the joint models intrinsically allowed for informative dropout due to death, they did not consider the issue of intermittent missing data for SLD and assumed that any missing values were missing completely at random (MCAR). This assumption could have been relaxed by imputing values for the missing data at each iteration, conditional upon covariates in the model, and assuming they were Missing at Random (MAR) (23). While we used non-informative prior distributions for most parameters in the model, a weakly informative Half-Normal ($0.5^2$) prior distribution was used for between-tumour standard deviation to help with model convergence. The exchangeable joint model was quite sensitive to the choice of prior due to the lack of information within each tumour subgroup, though this should be less of an issue for larger datasets.

For the longitudinal submodel a more flexible regression model could have been considered, though a quadratic form was initially considered but did not provide a better fit. Potentially the use of fractional polynomials may warrant consideration (24). For the survival submodel, a Weibull proportional hazards model was assumed, and whilst this appeared to be an appropriate model for the data (19), other potentially more flexible parametric models could be considered (24), including a range of parametric models within a Bayesian Model Averaging framework (25). The results presented are not adjusted for other covariates apart from tumour type, due to the relatively small number of events, but with additional follow-up these could be included in both the longitudinal and survival submodels.

For the exchangeable joint model, we assumed that the association between OS and SLD was exchangeable across all tumour sites – which may not be a reasonable assumption given the nature of the different tumour sites. Clinical expert opinion elicited that the association between OS and SLD could be exchangeable across some tumour sites, e.g. lung and thyroid, but not others. Given a larger dataset with more events, a partially exchangeable model may provide a better fit to the data and be deemed more clinically plausible (26).

Comparison of the overall or tumour site specific extrapolated summary measures, e.g. RMST, from the joint modelling approach with those from the more recent data-cut was undertaken at the population level. This could have been undertaken at the patient-level, but the methods for undertaking such calibration are currently under-developed (27). Furthermore, 125 patients were still censored at the later datacut, so follow-up data was not mature enough to enable the full validation of earlier survival predictions of individual patients.

Finally, we have considered SLD as a potentially predictive biomarker to aid prediction of OS and thus extrapolation of the study results. However, other biomarkers could also be considered, for example, health-related quality of life (HRQoL) data (in the case of this study EQ-5D). When we included EQ-5D by extending the joint models to include a multivariate formulation of the longitudinal submodel (5), its association with OS was not statistically significant, hence this case study did not warrant such an approach. However, it is likely that SLD and HRQoL have an inverse relationship, i.e. as SLD increases we would expect HRQoL to decrease (regardless of the relationship of either with OS). By adopting a Bayesian joint model, genuine informative prior distributions on the relationship between the biomarkers over time - which we might expect to have background information on - could be used.

## Conclusions

Joint modelling using intermediate outcomes such as tumour burden can offer an alternative approach to traditional survival modelling and may improve survival predictions from limited follow-up data. Although this case study was based on a relatively small and heterogeneous dataset, potential uses and benefits of adopting this approach were demonstrated, including the ability to model complex hierarchical data structures such as patients nested within tumour types. The joint modelling approach can also be extended to incorporate multiple longitudinal biomarkers in a multivariate modelling framework.

## Acknowledgements

The authors would like to thank Dr Sam Brilleman, Dr Michael Crowther and Prof Dimitris Rizopoulos for helpful software related discussions.

**Tables and figures**

**Figure 1: Sum of diameters of target lesions from treatment initiation by mortality status**

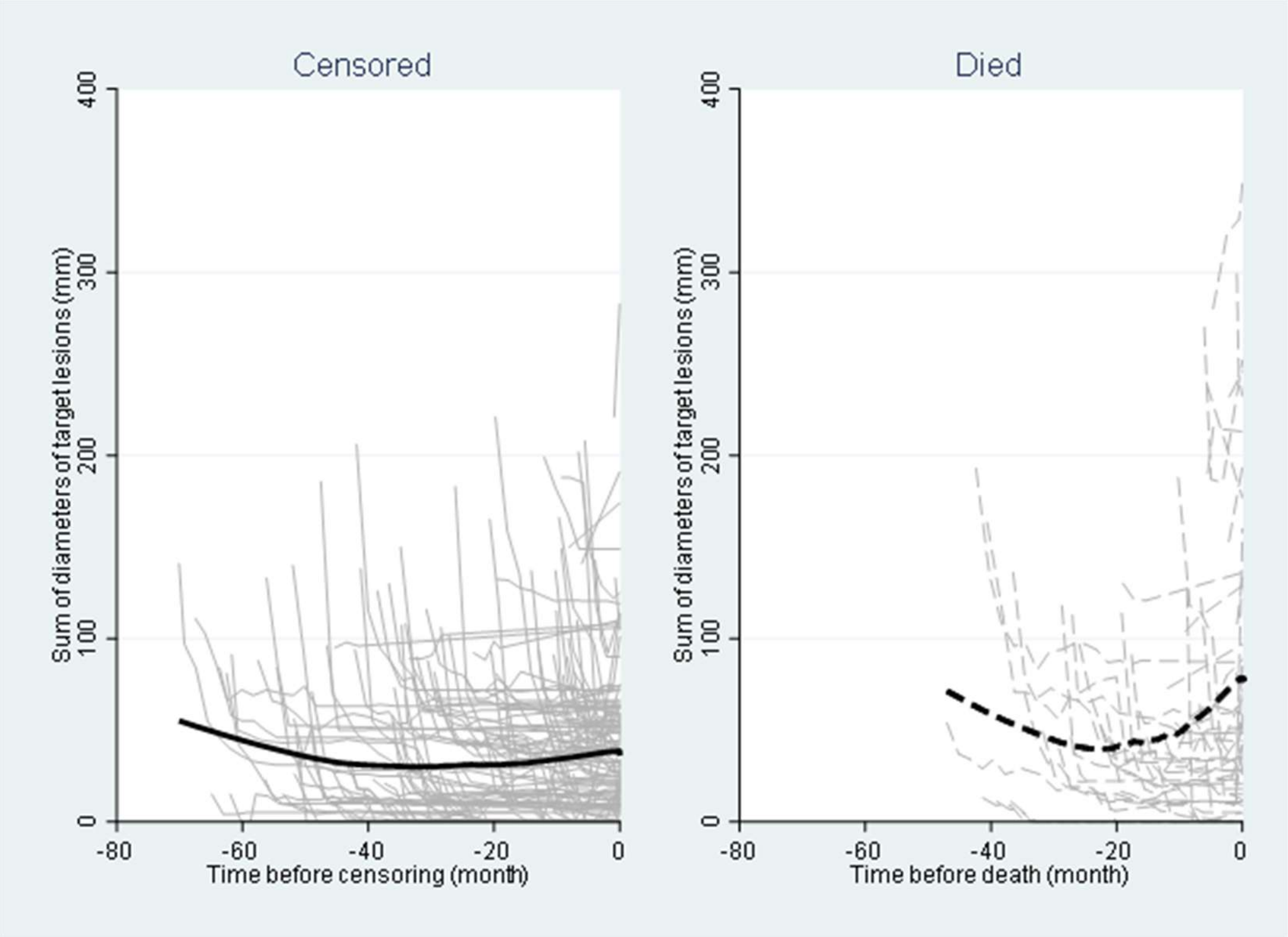

**Table 1: Association parameters and model fit for joint models of overall survival and sum of diameters of target lesions using a current value association structure; common, exchangeable, and independent across all tumour sites.**

| Model | Tumour type | Number of events/patients | Association hazard ratio[1] | 95% CrI | Model fit (DIC) |
|---|---|---|---|---|---|
| Common | All tumours | 58/196 | 1.09 | 1.05 to 1.14 | 16265 |
| Exchangeable | Soft tissue sarcoma | 13/65 | 1.10 | 1.00 to 1.22 | 15974 |
| | Thyroid | 9/30 | 1.08 | 1.00 to 1.18 | |
| | Salivary gland | 6/25 | 1.05 | 0.83 to 1.18 | |
| | Lung | 5/23 | 1.01 | 1.03 to 1.20 | |
| | Other | 25/53 | 1.08 | 1.03 to 1.14 | |
| Independent | Soft tissue sarcoma | 13/65 | 1.12 | 0.94 to 1.27 | 16018 |
| | Thyroid | 9/30 | 1.08 | 0.87 to 1.23 | |
| | Salivary gland | 6/25 | 0.78 | 0.59 to 1.09 | |
| | Lung | 5/23 | 1.15 | 1.03 to 1.27 | |
| | Other | 25/53 | 1.08 | 1.02 to 1.14 | |

[1] Association hazard ratio is the risk in mortality per 10mm increase in tumour burden.
CrI: credible interval, DIC: deviance information criterion.

**Table 2: Extrapolated survival results from Weibull proportional hazards model and posterior means from joint models of overall survival and sum of diameters of target lesions using a current value association structure with an assumption of common, exchangeable, or independent association across all tumour sites**

| Tumour site (events/patients) | Model | RMST over lifespan (years), 95% CI/Crl | Median over lifespan (years), 95% CI/CrI | Landmark survival at 10 years (%), 95% CI/CrI |
|---|---|---|---|---|
| All patients (58/196) | | | | |
| | Weibull | 8.04 (4.88 to 13.44) | 4.91 (3.25 to 6.57) | 27.7 (13.7 to 39.8) |
| | Common | 8.96 (5.11 to 13.84) | 4.63 (3.46 to 5.96) | 27.5 (16.3 to 38.8) |
| | Exchangeable | 9.35 (5.59 to 14.58) | 4.65 (3.57 to 5.98) | 28.1 (17.3 to 38.3) |
| | Independent | 12.17 (7.22 to 17.80) | 4.97 (3.70 to 6.40) | 32.2 (21.4 to 40.3) |
| Soft tissue sarcoma (13/65) | | | | |
| | Weibull | 12.05 (3.86 to 29.47) | 7.75 (5.07 to 10.42) | 41.7 (5.9 to 61.5) |
| | Common | 11.95 (5.74 to 21.13) | 7.46 (4.14 to 12.08) | 38.2 (21.5 to 58.5) |
| | Exchangeable | 13.36 (5.87 to 24.18) | 8.01 (4.24 to 13.40) | 40.6 (21.5 to 58.5) |
| | Independent | 19.43 (7.86 to 32.15) | 10.74 (4.88 to 18.56) | 49.0 (30.8 to 63.1) |
| Lung (5/23) | | | | |
| | Weibull | 4.53 (2.10 to 11.94) | 3.89 (1.21 to 6.56) | 6.5 (0.0 to 41.2) |
| | Common | 9.45 (3.12 to 19.61) | 6.27 (2.36 to 12.87) | 29.7 (4.3 to 56.5) |
| | Exchangeable | 9.89 (2.98 to 22.73) | 6.45 (2.27 to 14.03) | 29.9 (0.0 to 56.5) |
| | Independent | 9.63 (3.86 to 17.38) | 5.92 (2.77 to 10.71) | 30.6 (8.7 to 52.2) |
| Salivary gland (6/25) | | | | |
| | Weibull | 8.48 (3.10 to 25.14) | 6.81 (4.13 to 9.48) | 32.5 (0.0 to 60.7) |
| | Common | 13.75 (5.64 to 24.46) | 9.54 (4.55 to 17.08) | 44.0 (16.0 to 64.0) |
| | Exchangeable | 12.59 (5.77 to 22.16) | 8.81 (4.42 to 15.03) | 42.0 (20.0 to 64.0) |
| | Independent | 17.57 (6.48 to 30.81) | 9.94 (4.11 to 19.17) | 45.5 (20.0 to 64.0) |
| Thyroid (9/30) | | | | |
| | Weibull | 7.63 (2.76 to 23.79) | 5.15 (2.47 to 7.82) | 26.8 (0.15 to 52.0) |
| | Common | 8.21 (3.83 to 14.83) | 5.40 (3.19 to 8.81) | 26.5 (3.3 to 46.7) |
| | Exchangeable | 8.38 (3.97 to 14.49) | 5.44 (3.26 to 8.67) | 27.2 (10.0 to 50.0) |
| | Independent | 9.15 (4.29 to 15.20) | 5.58 (3.27 to 8.80) | 29.4 (10.0 to 50.0) |
| Other (25/53) | | | | |
| | Weibull | 3.46 (1.99 to 6.97) | 1.97 (0.00 to 4.64) | 7.4 (0.4 to 20.6) |
| | Common | 3.27 (2.13 to 4.73) | 2.05 (1.47 to 2.56) | 6.2 (0.0 to 15.1) |
| | Exchangeable | 3.24 (2.09 to 4.60) | 2.05 (1.49 to 2.56) | 6.0 (0.0 to 13.2) |
| | Independent | 3.52 (2.11 to 5.21) | 2.11 (1.46 to 2.56) | 7.5 (0.0 to 17.0) |

CI: confidence interval (for Weibull model), CrI: credible interval (for joint models), RMST: restricted mean survival time

**Figure 2: Extrapolated survival results from Weibull proportional hazards model and exchangeable joint model of overall survival and sum of diameters of target lesions using a current value association structure**

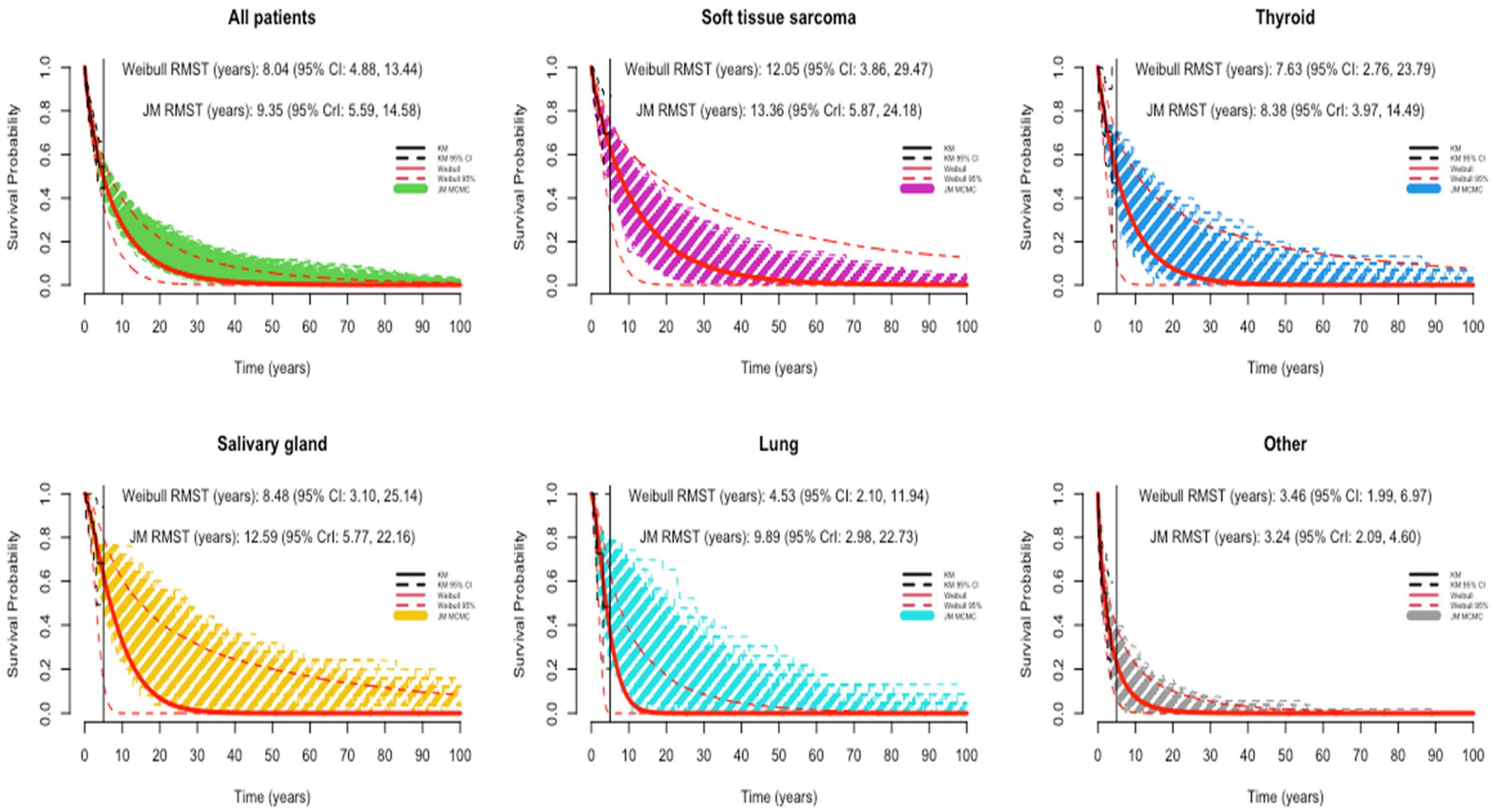

**Table 3: Extrapolated survival results from the exchangeable joint model of overall survival and sum of diameters of target lesions compared to the July 2023 datacut**

| Tumour site (events/patients) | Model | RMST at 5 years (years), 95% CI/CrI |
|---|---|---|
| All patients (71/196) | | |
| | Observed | 3.73 (3.51 to 3.95) |
| | Weibull | 3.50 (3.18 to 3.77) |
| | Exchangeable joint model | 3.41 (3.24 to 3.56) |
| Soft tissue sarcoma (19/65) | | |
| | Observed | 3.83 (3.49 to 4.18) |
| | Weibull | 3.96 (3.25 to 4.34) |
| | Exchangeable joint model | 3.84 (3.57 to 4.10) |
| Lung (9/23) | | |
| | Observed | 3.38 (2.67 to 4.08) |
| | Weibull | 3.47 (2.12 to 4.18) |
| | Exchangeable joint model | 3.67 (3.13 to 4.22) |
| Salivary gland (8/25) | | |
| | Observed | 4.35 (3.95 to 4.74) |
| | Weibull | 4.10 (3.05 to 4.58) |
| | Exchangeable joint model | 4.11 (3.90 to 4.26) |
| Thyroid (10/30) | | |
| | Observed | 4.02 (3.50 to 4.55) |
| | Weibull | 3.62 (2.56 to 4.20) |
| | Exchangeable joint model | 3.58 (3.28 to 3.86) |
| Other (25/53) | | |
| | Observed | 3.12 (2.62 to 3.63) |
| | Weibull | 2.41 (1.69 to 2.97) |
| | Exchangeable joint model | 2.34 (2.03 to 2.63) |

CI: confidence interval (for Weibull model), CrI: credible intervals (for joint models), RMST: restricted mean survival time

**Supplementary appendix**

**Figure S1: Kaplan-Meier curve for overall survival**

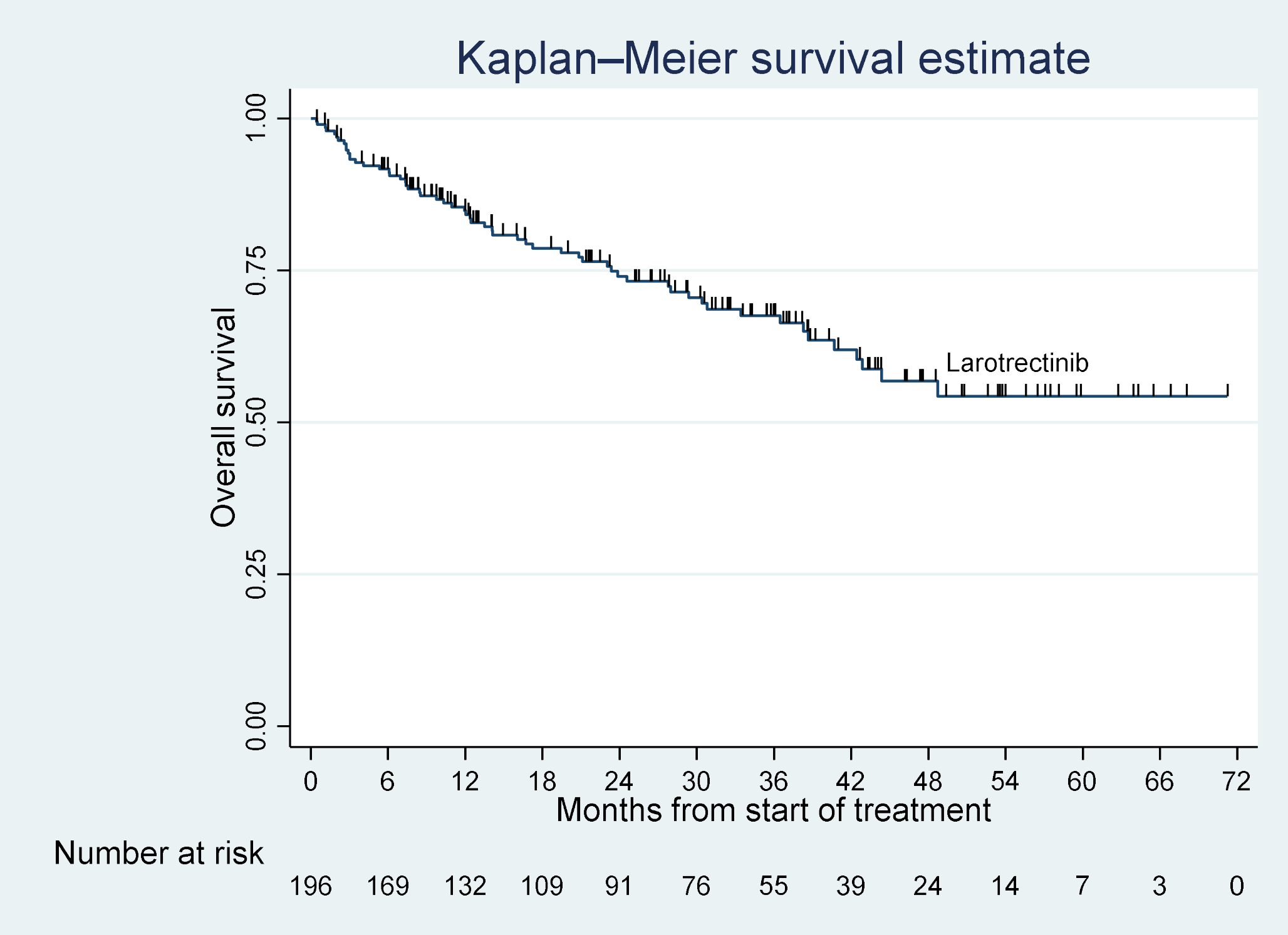

**Table S1: Baseline clinical characteristics and number of mortality events**

| | **Number (%) of mortality events (n=196)** |
|---|---|
| Age group (years): | |
| <18 | 4/40 (10.0) |
| 18 to <45 | 13/46 (28.3) |
| 45 to <65 | 24/68 (35.3) |
| ≥65 | 17/42 (40.5) |
| Median age {IQR} | 47 {25.5, 63} |
| ECOG performance status: | |
| 0 | 10/81 (12.4) |
| 1 | 32/85 (37.6) |
| 2 | 13/24 (54.2) |
| 3 | 3/6 (50.0) |
| Known metastasis at time of larotrectinib initiation: | |
| Locally advanced | 2/29 (6.9) |
| Metastatic | 56/167 (33.5) |
| Site of primary tumour: | |
| Soft tissue sarcoma | 13/65 (20.0) |
| Thyroid | 9/30 (30.0) |
| Salivary gland | 6/25 (24.0) |
| Lung | 5/23 (21.7) |
| Other | 25/53 (47.2) |
| Total | 58/196 (29.6) |

IQR: interquartile range, ECOG: Eastern Cooperative Oncology Group